\documentclass[acmtog,screen,nonacm]{acmart}
\renewcommand\footnotetextcopyrightpermission[1]{} 
\usepackage{booktabs} 

\usepackage[ruled]{algorithm2e} 

\SetAlFnt{\small}
\SetAlCapFnt{\small}
\SetAlCapNameFnt{\small}
\SetAlCapHSkip{0pt}

\acmJournal{TOG}

\begin{document}
\title{A New Split Algorithm for 3D Gaussian Splatting}

\author{Qiyuan Feng}
\affiliation{%
  \institution{BNRist, Department of Computer Science and Technology, Tsinghua University}
  \country{China}}
\author{Gengchen Cao}
\affiliation{%
  \institution{BNRist, Department of Computer Science and Technology, Tsinghua University}
  \country{China}}
\author{Haoxiang Chen}
\affiliation{%
  \institution{BNRist, Department of Computer Science and Technology, Tsinghua University}
  \country{China}}
\author{Tai-Jiang Mu}
\affiliation{%
  \institution{BNRist, Department of Computer Science and Technology, Tsinghua University}
  \country{China}}
\author{Ralph R. Martin}
\affiliation{%
  \institution{Cardiff University}
  \country{UK}}
\author{Shi-Min Hu}
\affiliation{%
  \institution{BNRist, Department of Computer Science and Technology, Tsinghua University}
  \country{China}}

\renewcommand\shortauthors{Feng, Q. et al}


%
%


%
%


\begin{teaserfigure}
  \includegraphics[width=\textwidth]{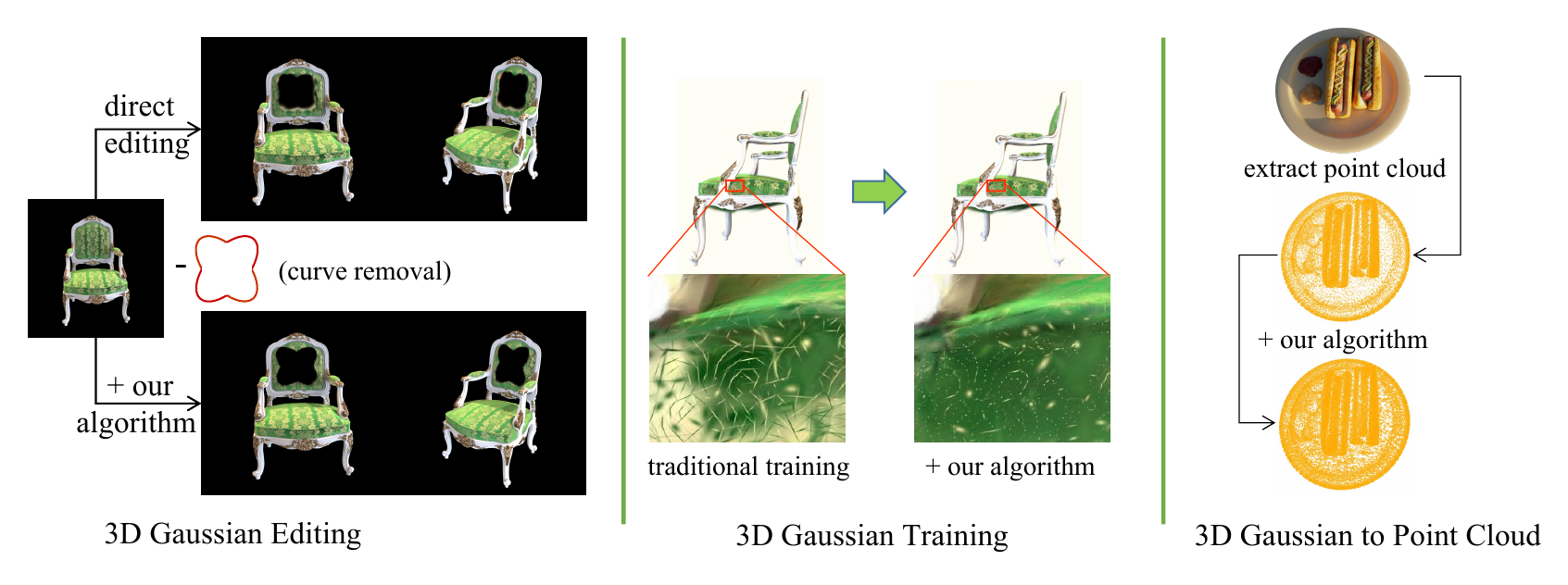}
  \caption{Benefiting from our new Gaussian splitting algorithm, we can produce more uniform, surface-bounded 3D Gaussian splatting models, enabling their explicit editing to produce clearer surfaces and the extraction of denser point clouds for untextured regions. Our splitting method also benefits 3D Gaussian learning, rendering views of higher quality.}
  \Description{My teaser}
  \label{fig:teaser}
\end{teaserfigure}

\begin{abstract}
  3D Gaussian splatting models, as a novel explicit 3D representation, have been applied in many domains recently, such as explicit geometric editing and geometry generation. Progress has been rapid.  However, due to their mixed scales and cluttered shapes, 3D Gaussian splatting models can produce a blurred or needle-like effect near the surface. At the same time, 3D Gaussian splatting models tend to flatten large untextured regions, yielding a very sparse point cloud. 
  These problems are caused by the non-uniform nature of 3D Gaussian splatting models, so in this paper, we propose a new 3D Gaussian splitting algorithm, which can produce a more uniform and surface-bounded 3D Gaussian splatting model. Our algorithm splits an $N$-dimensional Gaussian into two $N$-dimensional Gaussians. It ensures consistency of mathematical characteristics and similarity of appearance, allowing resulting 3D Gaussian splatting models to be more uniform and a better fit to the underlying surface, and thus more suitable for explicit editing, point cloud extraction and other tasks. Meanwhile, our 3D Gaussian splitting approach has a very simple closed-form solution, making it readily applicable to any 3D Gaussian model. 
\end{abstract}

\maketitle

\section{introduction}
3D representation is a key part of artistic model design, special effects production, and VR/AR applications.
Some implicit 3D representations like neural radiance fields (NeRFs)~\cite{mildenhall2021nerf}, have achieved great results in rendering quality, but it is very difficult to edit them and to provide real-time rendering. 
Instead, most games and modeling use explicit representations, especially point clouds and meshes, which have faster rendering speed and are easy to edit. 
3D Gaussian splatting (3D GS)~\cite{kerbl20233d}, as a novel 3D explicit representation, uses a series of 3D positions, opacity, anisotropic covariance, and spherical harmonic (SH) coefficients to represent a 3D scene. Gaussian splatting models can be rapidly rendered using a parallel differentiable rasterization approach. 
This novel 3D representation has been widely used for 3D editing~\cite{fang2023gaussianeditor,chen2023gaussianeditor}, physical simulation~\cite{xie2023physgaussian} and 3D geometry generation~\cite{tang2023dreamgaussian}. 

While 3D GS has many advantages compared to other implicit 3D representations, it suffers from two issues: (i) \emph{scale inhomogeneity}: some Gaussians have much greater value in every principal component of covariance than others, and (ii) \emph{structural inhomogeneity}: some Gaussians have much greater value in some principal components of covariance than other principal components. 
As a consequence, during 3D Gaussian splatting model editing and rendering, this inhomogeneity of 3D Gaussians  makes the boundaries of objects appear blurred and untidy,
as shown in the direct editing and traditional training results in Fig.~\ref{fig:teaser}.
Accumulated editing operations can lead to the collapse of the whole scene. Furthermore, point clouds extracted from 3D GS can be very sparse for untextured regions due to the scale inhomogeneity. This sparseness can cause difficulties in downstream tasks, such as normal estimation~\cite{xu2023globally,dey2005normal,li2010robust} and point cloud segmentation~\cite{qi2017pointnet,li2018pointcnn,guo2021pct}.

To solve the problem of scale inhomogeneity and structural inhomogeneity, we should make the inhomogeneous principal components more uniform, yet retain 3D visual consistency where affected. 
We do so by splitting Gaussians with undesirable shapes.
Inspired by Hall et al.~\shortcite{hall2002adding}, who proposed an algorithm to add and subtract eigenspaces by eigenvalue decomposition and singular value decomposition, we propose an algorithm to decompose the eigenspace into two independent eigenspaces while preserving its mathematical characteristics. 
For a 3D GS model, its mathematical characteristics are the parameters affecting visual effects: opacity, 3D position, and anisotropic covariance. 
Preserving mathematical characteristics during Gaussian splitting can be achieved by conserving zero-, first-, and second-moments. 
Conserving these allows us to model Gaussian splitting via a set of integral tensor equations, which have a concise and efficient closed-form solution.
This allows our splitting algorithm to be used with any 3D Gaussian model.
We also propose a metric to measure how accurately the 3D Gaussian fits the underlying scene after splitting. 

We show how our algorithm can benefit various applications. For Gaussian-based explicit editing, our method can produce clearer boundaries. For point cloud extraction from 3D Gaussians, our method represents the large flattened, texture-less regions with more uniform Gaussians, thus resulting in a denser point cloud. Our method also benefits  3D Gaussian learning, rendering novel views of higher quality.

In summary, this paper makes the following contributions:
\begin{itemize}
    \item We give a new splitting algorithm for 3D Gaussians, which can be used to produce a more uniform 3D Gaussian splatting model.
    \item We show the proposed Gaussian splitting approach has a closed-form solution, which can be efficient and is thus usable for any 3D Gaussian model.
    \item We demonstrate that our Gaussian splitting can benefit various applications, including explicit editing, point cloud densification, and 3D Gaussian learning.
\end{itemize}

\section{related works}
\subsection{3D representations}

A 3D representation is a data structure for handling and storing 3D geometry. Depending on the requirements, there are many different kinds of 3D representations; they can be divided into explicit and implicit representations.
Implicit representations include NeRF~\cite{mildenhall2021nerf}, NeuS~\cite{wang2021neus}, iso-surface~\cite{3dshape2vecset}, etc.
Such implicit representations, together with neural networks, can render views of high quality and are more capable of modeling scenes from other modalities such as text, and images~\cite{poole2022dreamfusion, wang2023prolificdreamer}. However, they are very difficult to edit because of their implicit nature. 
Works committed to solving this problem include the following. Dmtet~\cite{shen2021deep} can extract a differentiable iso-surface from a 3D implicit representation. NeRFEditor~\cite{sun2022nerfeditor} applied StyleGAN~\cite{karras2019style} to edit a NeRF.
DE-NeRF~\cite{wu2023nerf} decomposed a NeRF to accomplish editing such as changing lighting, materials, etc. However, these editing methods for implicit representations usually involve very complicated operations and are inefficient.

On the contrary, explicit representations provide a more direct 3D geometric structure, so editing operations are quite intuitive for creators and designers. A 3D mesh can be modified by an as-rigid-as-possible algorithm~\cite{sorkine2007rigid}, and also can be refined by loop subdivision~\cite{loop1987smooth}. Point clouds can be readily simulated~\cite{chen2020moving}
 and segmented~\cite{qi2017pointnet,guo2021pct}. However, an explicit representation is very tedious to create, requiring much work by human artists.

\subsection{3D Gaussian Splatting Models and  Applications}

3D Gaussian splatting~\cite{kerbl20233d} provides a novel explicit 3D  representation, which is convenient to obtain from images with deep learning techniques. Its explicit nature also makes it more suitable for real-time rendering than implicit representations.

Therefore, 3D Gaussian splatting has a wide range of applications. Several works have employed it in simultaneous localization and mapping (SLAM)~\cite{matsuki2023gaussian}.
Huang et al.~\shortcite{huang2023photo} has utilized their fast training speed to achieve better results in monocular vision. 
Some dynamic 3D radiance field tasks have been migrated to 3D Gaussian splatting modeling. For example, Wu et al.~\shortcite{wu20234d} combined 3D Gaussian splatting with a tri-plane approach~\cite{chen2022tensorf} to achieve better results for 4D scene reconstruction. For AI-based content generation, Chen et al.~\shortcite{chen2023text} utilized diffusion models~\cite{ho2020denoising, song2020denoising, lu2022dpm} to guide the generation of 3D Gaussian splatting models. Avatar generation with 3D Gaussian splatting~\cite{zielonka2023drivable} is also a popular research area. Works like~\cite{chen2023gaussianeditor, tang2023dreamgaussian} utilize diffusion to accomplish editing.

But as an explicit representation, how to edit a 3D Gaussian splatting model without extra training is very important. Real-time interactive performance is crucial for editing, but is almost impossible for deep neural network-based editing. In addition,  mutual transformation between Gaussian splatting and other explicit representations such as point clouds and meshes, is also important, to make  3D Gaussian splatting models compatible with current geometry processing pipelines aimed at explicit representations.





\section{Method}
Our goal is to replace the inhomogeneous Gaussians with more uniform Gaussians. We first briefly explain the 3D Gaussian representation in Sec.~\ref{sec:intro3DGaussian}. Then we model the splitting problem in Sec.~\ref{sec:problem}. We introduce some conservation relationships used in the splitting process and explain their physical meaning in Sec.~\ref{sec:cons}. A derivation of and solution to the splitting problem are given in Sec.~\ref{sec:derive}. Finally, we consider implementation details in Sec.~\ref{sec:abnormal}. 

\subsection{3D Gaussian Representation}
\label{sec:intro3DGaussian}

A 3D Gaussian splatting model 
represents a 3D target using a set of 3D Gaussians, each of which is characterized by its 3D position $\mathbf{\mu}$, opacity $\alpha$, covariance $\Sigma$, and color, with spherical harmonic (SH) coefficients.
A 3D point with a coordinate $\mathbf{x}$ covered by a 3D Gaussian $k$ obtains its opacity $\alpha(\mathbf{x})$ from that Gaussian (if not specifically stated, the opacity $\alpha$ in this paper represents the part except for the Gaussian distribution normalization coefficient):
\begin{align}
    \alpha(\mathbf{x}) &= \alpha_k\ \textrm{pdf}_k(\mathbf{x}) \label{1a}\\
    \textrm{pdf}_k(\mathbf{x}) &= \frac{1}{(2\pi)^{3/2}|\Sigma_k|}\exp(-\frac{1}{2}(\mathbf{x}-\mathbf{\mu}_k)^T\Sigma_k^{-1}(\mathbf{x}-\mathbf{\mu}_k))\label{1b}
\end{align}
where $\rm{pdf}_k$ denotes the probability density function of Gaussian $k$. Due to the positive definiteness of covariance $\Sigma$, it can be decomposed as follows to  save storage space:
$$
\Sigma = RSS^TR^T
$$
where $R$ is the rotation matrix and $S$ is the principal direction scale diagonal matrix. $R$ can be stored as a quaternion $q$. All of these can be converted to their respective matrices and combined, making sure to normalize $q$ to obtain a valid unit quaternion.

{The original 3D Gaussian Splatting method directly duplicates a copy of the Gaussian with naive scaling and random translation when it needs to increase the number of Gaussians.}

\subsection{Problem Definition for Gaussian Splitting}
\label{sec:problem}

\begin{figure}[t]
    \centering
    \includegraphics[width=0.9\linewidth]{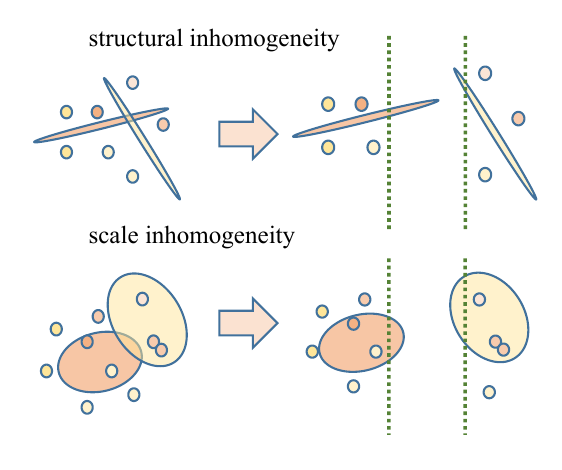}
    \caption{
    Structural (above) and scale (below) inhomogeneities (2D examples). If an editing operation is performed, here shown as cutting at a plane and pulling apart, the results will include parts crossing the plane:  blurring and needle-like bulges will occur.  Small circles represent more homogeneous Gaussians.
    }
    \label{fig:eliasplit-label}
\end{figure}

Since the covariance matrix $\Sigma$ is square and symmetric, it can be 
decomposed as:
$$
\Sigma = R\Lambda R^T
$$
where $\Lambda$ is a diagonal matrix, whose elements are the principal component's squared scales, 
and each column vector of $R$ represents a principal component's direction vector. Since the covariance matrix is a symmetric positive matrix, the three principal component direction vectors are mutually perpendicular unit vectors. 

The 3D Gaussian is completely central symmetric and axisymmetric.
Therefore, if all scales of a Gaussian are significantly larger than other Gaussians' scales (scale inhomogeneity), huge bulges will be caused when editing and huge voids will arise during point cloud extraction. If the scale of some principal component of a Gaussian is greater than other principal components (structural inhomogeneity),  there will be many needle-like objects near the surface, as demonstrated in Fig.~\ref{fig:eliasplit-label}.

At the same time, since explicit editing and point cloud extraction are carried out on the trained 3D Gaussian model and must usually be efficient for interaction, a  method to eliminate these inhomogeneities should not need further training. 
Moreover, if the visual consistency cannot be guaranteed before and after processing, {artifacts will accumulate during editing and extraction}.
{If we use a direct copy-based splitting strategy as the original Gaussian Splatting~\cite{kerbl20233d}} during the training split process, there will be an unstable loss decline, making the previous training invalid.

Our problem is how to shrink an inhomogeneous 3D Gaussian's principal components while maintaining its 3D visual consistency. 
Instead of shrinking, our approach is based on splitting, since any complex curved split can be approximately represented by splitting with many planes.

Therefore, the problem above is reduced to how to split a 3D Gaussian using a 3D plane while retaining its 3D visual consistency. A 3D plane can be expressed as:
\begin{align}
P(\mathbf{n},d)=\{\mathbf{x}\in \mathbb{R}^3 |\hat{P}(\mathbf{n}, d)(\mathbf{x}) \triangleq \mathbf{n}\cdot\mathbf{x} + d = 0 \}
\end{align}
where $\mathbf{n}$ denotes the unit normal of the plane, $d$ is its distance from the origin and $\hat{P}$ means directed distance. 
We then formally define the splitting of a Gaussian $(\alpha_0, \mathbf{\mu_0}, \Sigma_0)$  into two new Gaussians $(\alpha_l, \mathbf{\mu_l}, \Sigma_l)$ and $(\alpha_r, \mathbf{\mu_r}, \Sigma_r)$  at the splitting plane $P$ as:
$$
\alpha_l, \alpha_r, \mu_l,\mu_r,\Sigma_l,\Sigma_r = F(\alpha_0, \mu_0, \Sigma_0, P(\mathbf{n},d))
$$
which aims to minimize the intersection between the two new Gaussians while maximizing the similarity between each new Gaussian and their own part of the original:
\begin{align}
&\underset{\alpha_k,\Sigma_k,\mu_k}{\operatorname {argmin}}\int_{R^3\setminus\mathbb{V}_{k}}\alpha_k\textrm{pdf}_k(\mathbf{v})dV, k\in \{l,r\}\\
&\underset{\alpha_k,\Sigma_k,\mu_k}{\operatorname
{argmin}}|\int_{\mathbb{V}_{k}}\alpha_k\textrm{pdf}_k(\mathbf{v})dV-\int_{\mathbb{V}_{k}}\alpha_0\textrm{pdf}_0(\mathbf{v})dV|, k\in \{l,r\}
\end{align}
with:
$$
\begin{aligned}
\mathbb{V}_{l} &= \{\mathbf{x}\in \mathbb{R}^3|\hat{P}(\mathbf{n},d)(\mathbf{x})<0\}\\
\mathbb{V}_{r} &= \{\mathbf{x}\in \mathbb{R}^3|\hat{P}(\mathbf{n},d)(\mathbf{x})\geq 0\}
\end{aligned}
$$

The above goals also provide an evaluation metric for Gaussian splitting.
We simplify the optimization goal to get our splitting error metrics. This metric includes interval error and external excess {to evaluate 3D visual consistency and the extent to which blurring and needle-like objects exist beyond the boundary of editing, respectively}. Because the two new Gaussians and other Gaussians have unified characteristics, the interval error $E_i$ of the whole Gaussian model
 $\mathbb{G}$:
\begin{align}
E_i = \frac{1}{W}\sum_{g \in \mathbb{G}}{|\int_{\mathbb{V}_g}}\alpha_g\textrm{pdf}_g(\mathbf{v})dV - \int_{\mathbb{V}_g}\alpha_0\textrm{pdf}_0(\mathbf{v})dV|
\end{align}
where $W$ is the normalization coefficient, {defined as the number of all split Gaussians}, and $\mathbb{V}_g$ is the half-space where the 3D Gaussian $g$ is located.
To evaluate the cross-border situation of splitting, the external excess $E_e$ is defined as:
\begin{align}
E_e = \frac{1}{W}\sum_{g \in \mathbb{G}}{\int_{R^3\setminus\mathbb{V}_g}}\alpha_g\textrm{pdf}_g(\mathbf{v})dV
\end{align}

Lower $E_i$ and $E_e$ lead to more similar visual effects of the new Gaussians to those of the original Gaussian.

\begin{figure}[t]
    \centering
    \includegraphics[width=\linewidth]{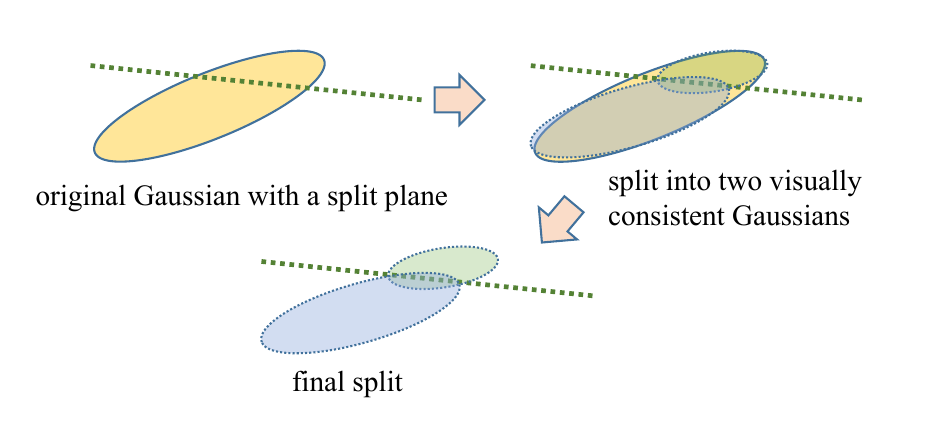}
    \caption{
    2D  illustration of the Gaussian splitting process. Three conservation rules are used to ensure visual consistency.}
    \label{fig:splitexample-label}
\end{figure}

\subsection{Conservation and Constraints}
\label{sec:cons}

In order to ensure the consistency of mathematical parameters as illustrated in Fig.~\ref{fig:splitexample-label}, 
we aim to conserve several quantities during splitting. Local weights of the probability density function are equivalent to opacity. Thus, to ensure that the new Gaussian looks the same as the original one from any perspective, the cumulative distribution of opacity should be conserved, i.e.,  zero-order moment:
\begin{align}
\int_{\mathbb{R}^3}\alpha_0\textrm{pdf}_0(\mathbf{v})dV = \int_{\mathbb{R}^3}\alpha_l\textrm{pdf}_l(\mathbf{v})dV + \int_{\mathbb{R}^3}\alpha_r\textrm{pdf}_r(\mathbf{v})dV
\label{zero conservation}
\end{align}

Similarly, in order to ensure that the location of the 3D Gaussian center $\mathbf{\mu}$ remains unchanged visually, we should conserve the first-order moment:
\begin{align}
\int_{\mathbb{R}^3}\alpha_0\mathbf{v}\textrm{pdf}_0(\mathbf{v})dV = \int_{\mathbb{R}^3}\alpha_l\mathbf{v}\textrm{pdf}_l(\mathbf{v})dV + \int_{\mathbb{R}^3}\alpha_r\mathbf{v}\textrm{pdf}_r(\mathbf{v})dV
\label{first conservation}
\end{align}

For each point, to ensure that the distribution is as consistent as possible, we hope that the mean of the outer product matrix will not change either, requiring the conservation of second-order moment:
\begin{align}
\int_{\mathbb{R}^3}\alpha_0\mathbf{v}\mathbf{v}^T\textrm{pdf}_0(\mathbf{v})dV = \int_{\mathbb{R}^3}\alpha_l\mathbf{v}\mathbf{v}^T\textrm{pdf}_l(\mathbf{v})dV + \int_{\mathbb{R}^3}\alpha_r\mathbf{v}\mathbf{v}^T\textrm{pdf}_r(\mathbf{v})dV
\label{second conservation}
\end{align}

 Via these conserved quantities, we can reduce the splitting problem to an optimization problem constrained by this integral tensor equation set (ITEs). This problem is difficult to solve in general, so we look for a special solution.
 
\subsection{Closed Form Solutions for Guassian Splitting}
\label{sec:derive}

{A special solution to the ITEs is the following integral problem. This solution separates the two Gaussians in spatial for a smaller $E_e$ and ensures that the visual consistency of the left and right sides separately for smaller $E_i$}:
\begin{align}
\label{eq:ITE-1}
\int_{\mathbb{V}_{k}}\alpha_0 \textrm{pdf}_0(\mathbf{v})dV &= \int_{\mathbb{R}^3}\alpha_{k}\textrm{pdf}_k(\mathbf{v})dV, k\in\{l,r\} \\
\label{eq:ITE-2}
\int_{\mathbb{V}_{k}}\alpha_0\mathbf{v}\textrm{pdf}_0(\mathbf{v})dV &= \int_{\mathbb{R}^3}\alpha_{k}\mathbf{v}\textrm{pdf}_k(\mathbf{v})dV, k\in\{l,r\} \\
\label{eq:ITE-3}
\int_{\mathbb{V}_{k}}\alpha_0\mathbf{v}\mathbf{v}^T\textrm{pdf}_0(\mathbf{v})dV &= \int_{\mathbb{R}^3}\alpha_{k}\mathbf{v}\mathbf{v}^T\textrm{pdf}_k(\mathbf{v})dV, k\in\{l,r\}
\end{align}

\begin{figure*}[!t]
    \centering
    \includegraphics[width=\linewidth]{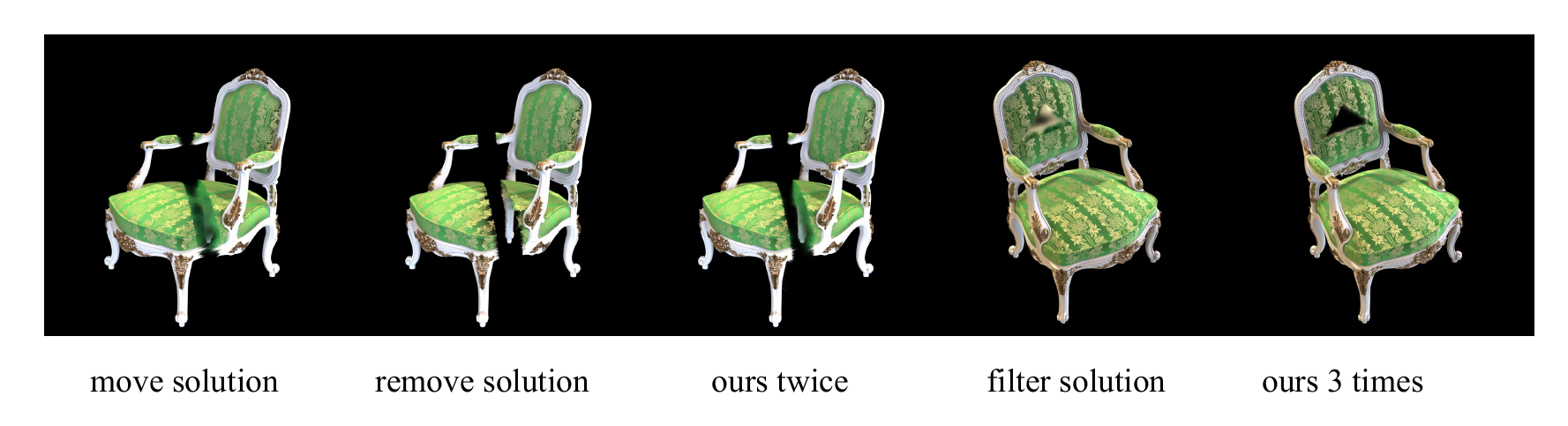}
    \caption{
    The first three examples split the chair into two pieces directly along a randomly chosen plane passing through the origin with $\mathbf{n} = [0.5401,0.8316,0.0963]^T$; the last two remove a triangular model from the chair's back. We compare our method to three other baselines: the move solution which moves any  Gaussians directly, the remove solution which removes any  Gaussians in the gap, and the filter solution which removes any Gaussian whose {position is inside the bounding box or the closed curve}. We execute our algorithm twice and 3 times in different situations for better results.
 {The move solution results visible components in the gap while the remove solution leads to many holes. The filter solution produces many artifacts. Our solution produces the cleanest boundaries near the cuts.} 
    }
    \label{fig:edition-label}
\end{figure*}

Given that any 3D Gaussian $i$ satisfies the following equation concerning its second central moment:
\begin{align}
\Sigma_i + \mu_i\mu_i^T = \int_{\mathbb{R}^3}\mathbf{v}\mathbf{v}^T\textrm{pdf}_i(\mathbf{v})dV,
\end{align}
combining it with  Eqs.~(\ref{eq:ITE-1}-\ref{eq:ITE-3}), we can derive the closed-form solutions below for the split  equations:

\begin{align}
\alpha_k &= \alpha_0 C_k, k\in\{l,r\}\\
\mathbf{\mu}_l &= \mathbf{\mu}_0 -\frac{\mathbf{L}_0D}{\tau C_l} \\
\mathbf{\mu}_r &= \mathbf{\mu}_0 + \frac{\mathbf{L}_0D}{\tau C_r}\\
\mathbf{\Sigma}_l &= \mathbf{\Sigma}_0 + \frac{\mathbf{L}_0\mathbf{L}_0^T}{\tau^2}(\frac{d_0D}{\tau C_l}-\frac{D^2}{C_l^2})\\
\mathbf{\Sigma}_r &= \mathbf{\Sigma}_0 -\frac{\mathbf{L}_0\mathbf{L}_0^T}{\tau^2}(\frac{d_0D}{\tau C_{r}}+\frac{D^2}{C_{r}^2})
\end{align}
where:
\begin{align}
C_{l} &= \frac{1}{2}(1-\textrm{erf}(\frac{d_0}{\sqrt{2}\tau}))\\
C_{r} &= \frac{1}{2}(1+\textrm{erf}(\frac{d_0}{\sqrt{2}\tau}))\\
D &= \frac{1}{\sqrt{2\pi}}\exp(-\frac{d_0^2}{2\tau^2})\\
\mathbf{L}_0 &= \mathbf{\Sigma}_0\mathbf{n} \\
\tau &= \sqrt{\mathbf{n}^T\mathbf{\Sigma}_0\mathbf{n}}\\
d_0 &= \hat{P}(\mathbf{n},d)(\mathbf{\mu}_0)
\end{align}

{In the above, $\textrm{erf}$ is the error function from probability theory.  $C_l$ and $C_r$ are the spatial weights of the target Gaussians, $D$ is the update amplitude of the split process,
 $\mathbf{L}_0$ and $\tau$ are the projections of different orders
 and $d_0$ is the distance from the Gaussian position to the splitting plane.}
Detailed derivations are given in the Supplementary Material.

\subsection{Implementation Details for Splitting}
\label{sec:abnormal}

Due to floating point precision issues and the properties of Gaussian distributions, errors will accumulate as splitting continues (e.g. when multiple edits are performed). We use the following approach to compensate for these problems.

Firstly, if the splitting plane ${P}(\mathbf{n},d)$ is too far from the position of the original Gaussian, an almost invalid splitting is to create a Gaussian that is very similar to the original and another Gaussian that is almost invisible because of its very small $\alpha$.
 And if the splitting is applied to every Gaussian, there will be a huge amount of calculation and thus is inefficient. 
 Inspired by the Gaussian filter's influence range threshold, i.e, $3\times \max(\mathbf{Tri(S)})$ (where $\mathbf{Tri}$ denotes the main diagonal vector), we choose a threshold $\eta$ for $|d_0|$:
\begin{align}
\eta = 3 \times \max((R \cdot \mathbf{n}) \odot \mathbf{Tri}(S))
\end{align}
If $|d_0| < \eta$, we will split the Gaussian with the plane; otherwise, we just leave the original Gaussian unchanged.

Secondly, the weight denominator $C_{l}$ and $C_{r}$ may be very small, causing floating point accuracy to be compromised. Thus we add an offset $\epsilon=10^{-20}$ to $C_{l},C_{r}$ and $D$:
\begin{align}
C_{l} &= \frac{1}{2}(1-\textrm{erf}(\frac{d_0}{\sqrt{2}\tau})+\epsilon)\\
C_{r} &= \frac{1}{2}(1+\textrm{erf}(\frac{d_0}{\sqrt{2}\tau})+\epsilon)\\
D &= \frac{1}{\sqrt{2\pi}}(\exp(-\frac{d_0^2}{2\tau^2})+\epsilon)
\end{align}

Thirdly,  floating point precision issues may cause the eigenvalue matrix to deviate slightly from a real positive definite matrix. In order to ensure a positive definite resulting eigenvalue matrix $\Lambda_{ti}$, we  remove any deviation of  negative definite orientation from our eigenvalue matrix $\Lambda_i$:
\begin{align}
\Lambda_{ti} = \textrm{ReLU}(\Lambda_i),i\in{\mathbb{G}}
\end{align}

Fourthly, in computer graphics, right-handed coordinate system is typically used. A  rotation matrix $R$ defining a left-handed space cannot be transformed into a right-handed coordinate system by a normalized quaternion (the improper rotation problem). As a  3D Gaussian is completely centrally symmetric, we can turn the left-handed rotation matrix $R_i$ into a right-handed rotation matrix $R_{ti}$:
\begin{align}
    R_{ti} = R_i\det(R_i),i\in{\mathbb{G}}
\end{align}

In summary, the operations above are aimed at reducing the amount of matrix calculation, eliminating floating point precision errors, and removing logical inconsistencies in data transformation. 

\begin{figure*}
    \centering
    \includegraphics[width=0.95\linewidth]{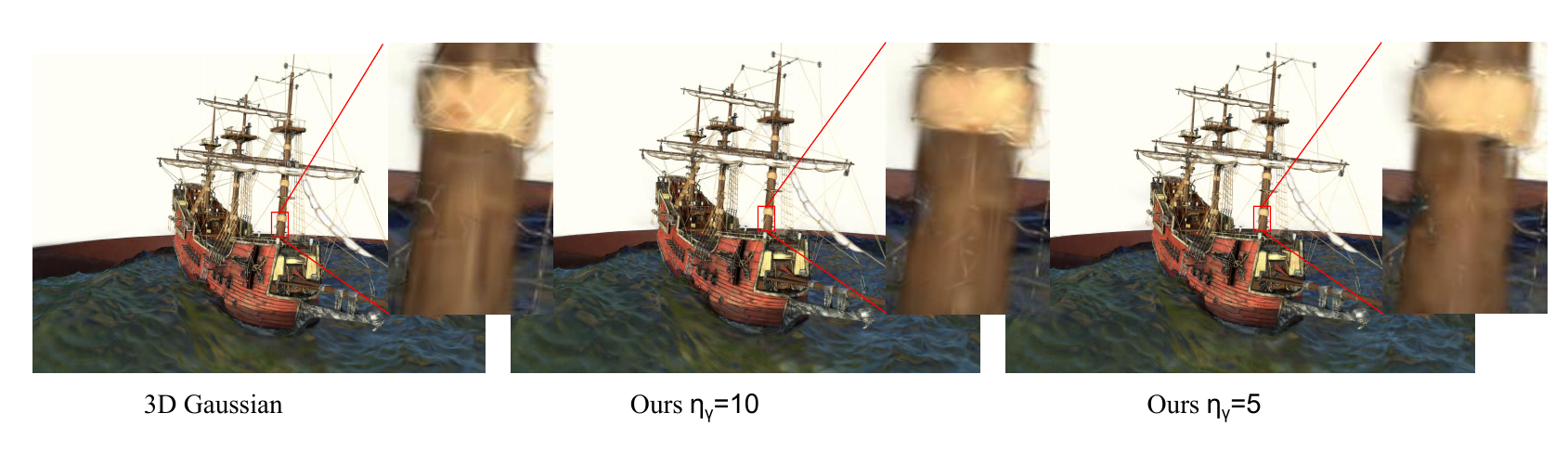}
    \caption{
    We integrate our splitting algorithm into the training of 3D Gaussian with different splitting thresholds $Th_{\gamma}$. Our method removes most of the inhomogeneous Gaussians and achieves better results.
    }
    \label{fig:training-label}
\end{figure*}

\section{Experiments}
To show the effectiveness of our splitting algorithm, we apply it to explicit Gaussian splatting model editing, 3D Gaussian learning and point cloud extraction from Gaussian splatting models, and compare our method with other solutions. 

\subsection{Editing by Splitting}
During explicit editing, it is common 
to split some parts of an object or remove some specific components. We consider three editing strategies.

The first one is to split an object by a flat 3D plane, denoted  ``plane split''. As explained in Sec.~\ref{sec:derive}, using a plane ${P}(\mathbf{n},d)$, we can split any inhomogeneous 3D Gaussian into more uniform ones and achieve better results as shown in the first three examples of Fig.~\ref{fig:edition-label}.
More results are shown in Fig.~\ref{figapp:planesplit-label}. 
Since our algorithm produces a Gaussian model after editing, it can be applied several times to achieve further splits.

The second one removes Gaussians inside a polygonal bounding box, denoted  ``polygon delete''. We have designed a simple case to remove Gaussians inside a triangular prism to test the effect of our algorithm. We can easily find the outward direction of every plane of a polygonal bounding box. Subproblems include determining whether a line segment or a 3D triangle and a Gaussian intersect. For a line segment, edge projection can be applied to determine whether  an intersection exists:
 with a Gaussian position $\mu_0$ and a line segment $\{\mathbf{A},\mathbf{B}\}$, define:
\begin{align}
    I = ((\mathbf{A}-\mathbf{\mu}_0)\cdot(\mathbf{A}-\mathbf{B}))((\mathbf{B}-\mathbf{\mu}_0)\cdot(\mathbf{A}-\mathbf{B}))
\end{align}
If $I \geq 0$ and the intersection threshold $\eta$ is guaranteed, the line segment and the Gaussian intersect. Obviously, a 3D triangle intersection can be found using the above test three times. The results are shown in the last example in Fig.~\ref{fig:edition-label}.
For a more complex polyhedral bounding box, we can use a combination of the above methods or ray casting algorithm to retain or remove the Gaussians in the bounding box.

The third one is to remove Gaussians inside a 3D closed curved surface $B(x,y,z) = 0$, denoted as ``curve delete''. We can compute the tangent normal $\mathbf{n}$ of the closest position on the curve to the Gaussian position as:
\begin{align}
\mathbf{n} = \frac{\nabla B}{||\nabla B||}
\end{align}
where $\mathbf{n}$ is an outward normal from the inside of the curve according to the definition of a 3D closed curve.
{Example results are show in Fig.~\ref{fig:teaser}}. More results are shown in Fig.~\ref{figapp:edit-label}.

\begin{table}
  \caption{$E_i$ and $E_e$ when editing the chair}
  \label{tab:edit on chair}
  \begin{tabular}{cccl}
    \toprule
    Situation&Editing Strategy&$E_i\downarrow$&$E_e\downarrow$\\
    \midrule
    plane split & move solution & 0.0000 & 0.5836\\
    plane split & remove solution & 3.4121 & 0.0000\\
    plane split & ours once & 0.5828 & 0.0009\\
    plane split & ours twice & 0.5830 & 0.0006\\
    polygon delete & filter solution & 0.0000 & 0.0668\\
    polygon delete & ours 3 times & 0.3610 & 0.0005\\
    curve delete & filter solution & 0.0000 & 0.0710\\
    curve delete & ours 3 times & 0.3027 & 0.0015\\
  \bottomrule
\end{tabular}
\end{table}

We evaluated $E_i$ and $E_e$ for the above three editing situations for different splitting methods using the chair object from the NeRF synthetic dataset~\cite{mildenhall2021nerf} and move the two pieces apart. Quantitative results are shown in Table~\ref{tab:edit on chair} and visual comparisons are presented in Fig.~\ref{fig:edition-label}. 
If both $E_i$ and $E_e$ are smaller, the results are better. 
However, due to the unimodal and symmetric shape of Gaussian distribution, it is impossible for both metrics to be smaller at the same time.
Although other methods can have a zero value for $E_i$ or $E_e$, there will be significant holes when $E_e = 0$, and significant blurring and needles when $E_i = 0$.
Our algorithm achieves a small value for both indicators and ensures minimal voids, blurring and needles.

\begin{table}[t]
  \caption{$E_i$ and $E_e$ plane splitting on Nerf synthetic dataset}
  \label{tab:whole split}
  \begin{tabular}{ccl}
    \toprule
    Editing Strategy&$E_i\downarrow$&$E_e\downarrow$\\
    \midrule
    move solution & 0.0000 & 2.3823\\
    remove solution & 15.2541 & 0.0000\\
    ours & 2.3819 & 0.0003\\
  \bottomrule
\end{tabular}
\end{table}

To evaluate the average performance on the whole NeRF synthetic dataset, we conducted an experiment by splitting each object with a plane, which passes through the center of the space with a randomly generated normal $\mathbf{n} = [0.5401,0.8316,0.0963]^T$. The results are shown in Table~\ref{tab:whole split}. Our method can greatly decrease the part beyond the split plane and ensure very small interval changes.

\subsection{Integration with 3D Gaussian Learning}
The 3D Gaussian splatting model has achieved high-quality rendering results with very fast training and rendering. 
But its results still suffer from blurring and spurs. There exist many needle-like bugles as shown in the first example in Fig.~\ref{fig:training-label} for the original 3D Gaussian splatting model. More results are shown in Fig.~\ref{figapp:zoom-in-label}. As we have explained in Sec.~\ref{sec:problem}, this is caused by inhomogeneous Gaussians. 

Since  structural inhomogeneity comes from differences in scale along different principal directions, we can compute the inhomogeneity  $\gamma$ of a Gaussian as the ratio between the largest and the second largest scales of the principal directions:
\begin{align}
pd_m &= \underset{i \in \{0,1,2\}}{\operatorname {\textrm{max}}}(\mathbf{Tri}[i]))\\
pd_s &= \underset{i \in \{0,1,2\}}{\operatorname {\textrm{second\_max}}}(\mathbf{Tri}[i]))\\
\gamma &= \frac{pd_m}{pd_s}
\end{align}

We can set a max threshold $\eta_{\gamma} = 5$ for $\gamma$; if  $\gamma$ of a Gaussian exceeds this threshold, the Gaussian needs to be split. During 3D Gaussian learning, the tangent normal is needed. We can assume that the normal direction is parallel to the principal direction of the max eigenvalues:
\begin{align}
\mathbf{n} = R[\underset{i \in \{0,1,2\}}{\operatorname {\textrm{argmax}}}(\mathbf{Tri}[i]))]
\end{align}
We set the split plane to pass through the center of the Gaussian, so we have a simplified version of our algorithm with $d_0 = 0$. Furthermore, $C_{l} = C_{r} = 0.5$ and $D = 1/{\sqrt{2\pi}}$. The first term of $\Sigma_l$ and $\Sigma_r$ is $0$. In this way, we can speed up the basic algorithm by saving calculations.

The training results for the ship object
 are shown in Fig.~\ref{fig:training-label}. Although the three views are the same from a distance, when viewed close up, the original simple 3D Gaussian has lots of needles. By incorporating our algorithm, the needles decrease greatly with a threshold of $\eta_{\gamma} = 10$ and hardly any needles result with a threshold of $\eta_{\gamma} = 5$.

We tested the threshold $\eta_\gamma$ every 5000 iterations starting from the 9999th iteration and ending at the 24999th iteration to determine whether to split or not. Since our splitting algorithm can be computed in parallel, it hardly increases the training time compared to the original 3D Gaussian splatting.
We also tested on the whole NeRF synthetic dataset with the same hyper-parameters as for the original 3D Gaussian splatting method. Quantitative results are presented in Table~\ref{tab:training results}, which gives Peak Signal-to-noise Ratio (PSNR), Structural Similarity (SSIM)~\cite{wang2004image}, and Learned Perceptual Image Patch Similarity (LPIPS)~\cite{zhang2018unreasonable}.
Our algorithm splits the inhomogeneous Gaussians, so that the trained Gaussian results have fewer needles, with better results for all three metrics.

\begin{table}
  \caption{3D Gaussian splatting training results with and without our splitting algorithm.}
  \label{tab:training results}
  \begin{tabular}{cccl}
    \toprule
    Method&PSNR$\uparrow$&SSIM$\uparrow$&LPIPS$\downarrow$\\
    \midrule
    3D GS~\cite{kerbl20233d} & 36.9356 & 0.9825 & 0.0236\\
    Ours & 37.1908 & 0.9839 & 0.0216\\
  \bottomrule
\end{tabular}
\end{table}

\subsection{Extracting Point Clouds from 3D Gaussian Splatting Models}

A 3D Gaussian Splatting model can be regarded as providing a kind of point cloud, in which each point represents a different shape. This characteristic makes the point cloud extracted from 3D Gaussians splatting models very uneven, with many gaps in large flat untextured regions, leading to severe problems for downstream tasks like normal extraction, point cloud segmentation, etc.

\begin{figure}[t]
    \centering
    \includegraphics[width=\linewidth]{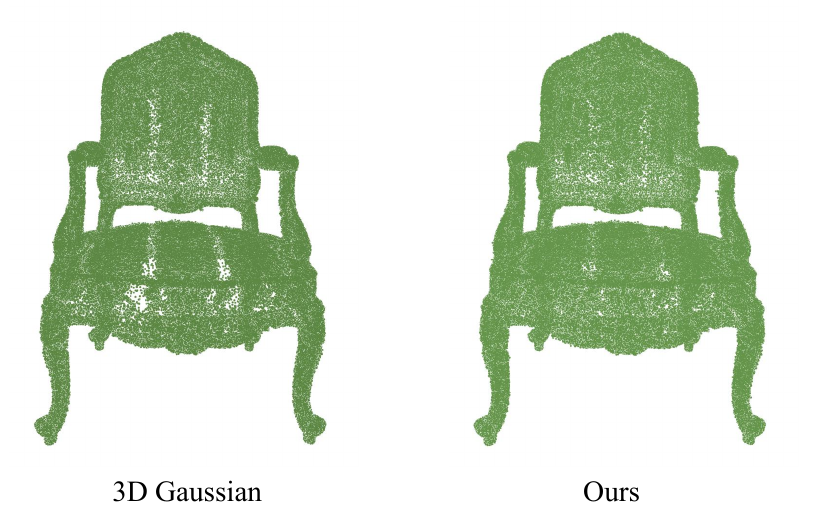}
    \caption{
    Our splitting algorithm helps to extract more uniform and denser point clouds from a well-trained 3D Gaussian model for the chair object.
    }
    \label{fig:refine-label}
\end{figure}

Because our algorithm can provide more homogeneous Gaussians, it can be used to increase the uniformity of extracted point clouds. Assume that the average of each Gaussian's maximum principal direction scale is $R_m$. We can define the degree of inhomogeneity between the $i$-th largest and the $j$-th largest eigenvalue ($i<j$) as:
\begin{align}
\gamma_{ij} = \frac{\mathbf{Tri}[i]}{\mathbf{Tri}[j]+R_m}, (i,j) \in \{(0,1), (1,2), (0,2)\}
\end{align}
If any $\gamma_{ij}$ is greater than the inhomogeneity threshold $\eta_{\gamma} = 2$, the Gaussian should be split at principal component $i$. Splitting at the center of a Gaussian will make the positions of new Gaussians too close. To ensure the new Gaussians' positions are far away from each other, we select the splitting parameters $d_0$ as $2\mathbf{Tri}[i]$ to ensure a proper interval between new Gaussians. 
An experimental result is shown in Fig.~\ref{fig:refine-label} and more results are shown in Fig.~\ref{figapp:refine-label}. Our method produces a more uniform and denser point cloud than the original 3D Gaussian Splatting model and has very few holes.

\section{Conclusion and Future Work}
In this paper, we model the problem of how to split an $N$-dimensional Gaussian into two independent $N$-dimensional Gaussians and present a closed-form solution for this problem. This enables our splitting algorithm to be readily used with any 3D Gaussian model processing and helps to produce more uniformly distributed Gaussian. In this way,  blurring and needle-like artifacts can be significantly reduced, benefiting various downstream applications. 

In the future, we hope to apply our work to conversions between 3D Gaussian Splatting models and other 3D representations, improvement of rendering quality, Gaussian splatting model editing, and some other geometric applications of Gaussian Splatting models. 
Moreover, there is an inverse to our method, to merge Gaussians, under the conservation constraints:
\begin{align}
\alpha_0 &= \alpha_l + \alpha_r\\
\mu_0 &= \frac{\alpha_l\mu_l+\alpha_r\mu_r}{\alpha_0}\\
\Sigma_0 &= \frac{\alpha_l\Sigma_l+\alpha_r\Sigma_r}{\alpha_0} + \frac{\alpha_l\mu_l\mu_l^T + \alpha_r\mu_r\mu_r^T}{\alpha_0} - \mu_0\mu_0^T
\end{align}
which can be used to greatly decrease storage requirements.

\bibliographystyle{ACM-Reference-Format}
\bibliography{sample-base}

\appendix

\begin{figure*}
    \centering
    \includegraphics[width=\linewidth]{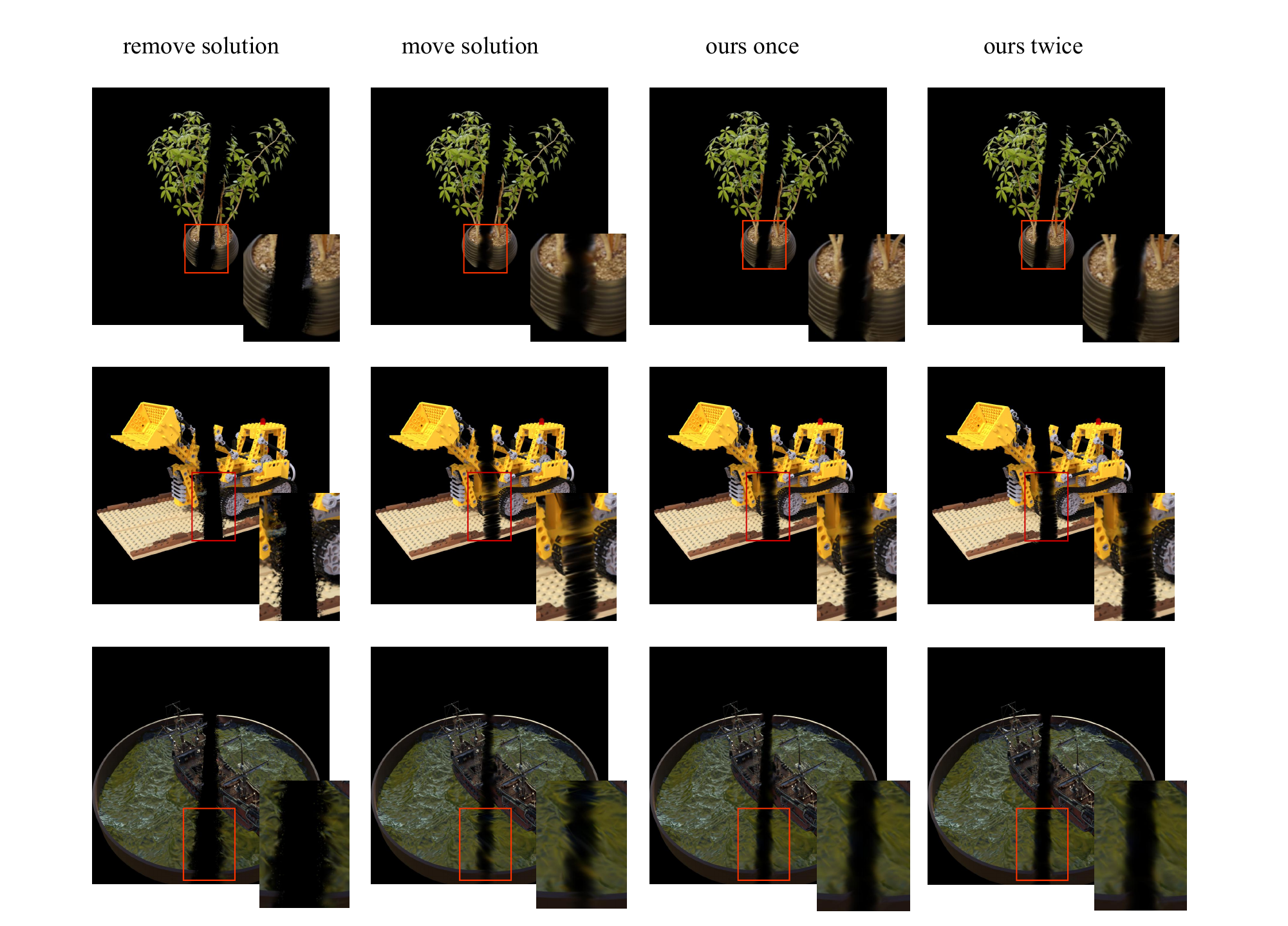}
    \caption{
    Various plane split results, which show the differences between splitting algorithms more clearly. The remove solution leads to more holes, and the move solution leads to more adhesion, while ours clearly separates the two parts with a clear boundary. 
    }
    \label{figapp:planesplit-label}
\end{figure*}

\begin{figure*}
    \centering
    \includegraphics[width=\linewidth]{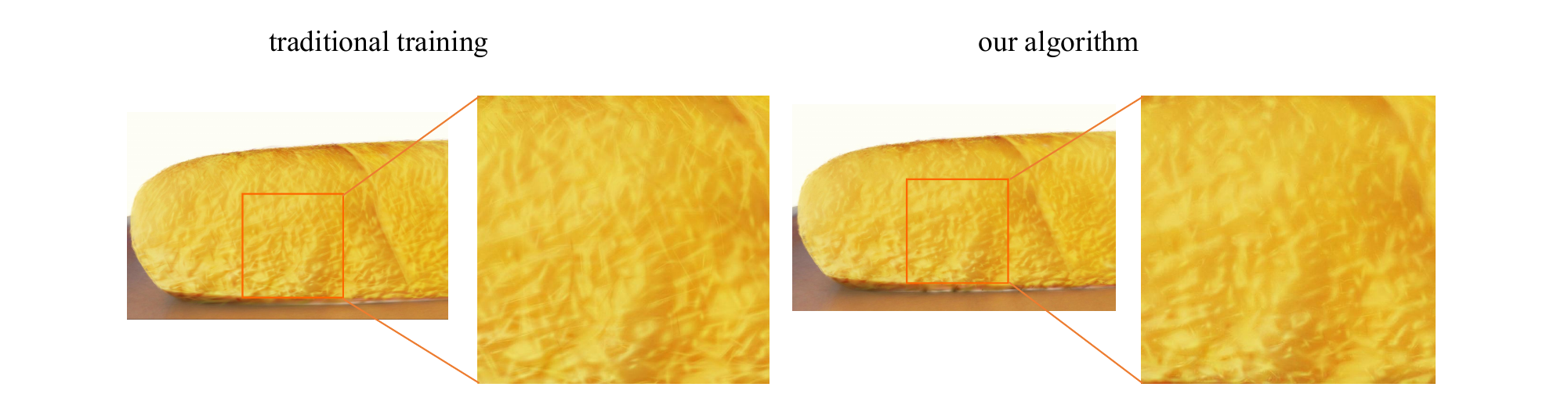}
    \caption{
     Incorporating our algorithm reduces blurring and needles compared to the standard 3D Gaussian splatting model training.
    }
    \label{figapp:zoom-in-label}
\end{figure*}

\begin{figure*}
    \centering
    \includegraphics[width=\linewidth]{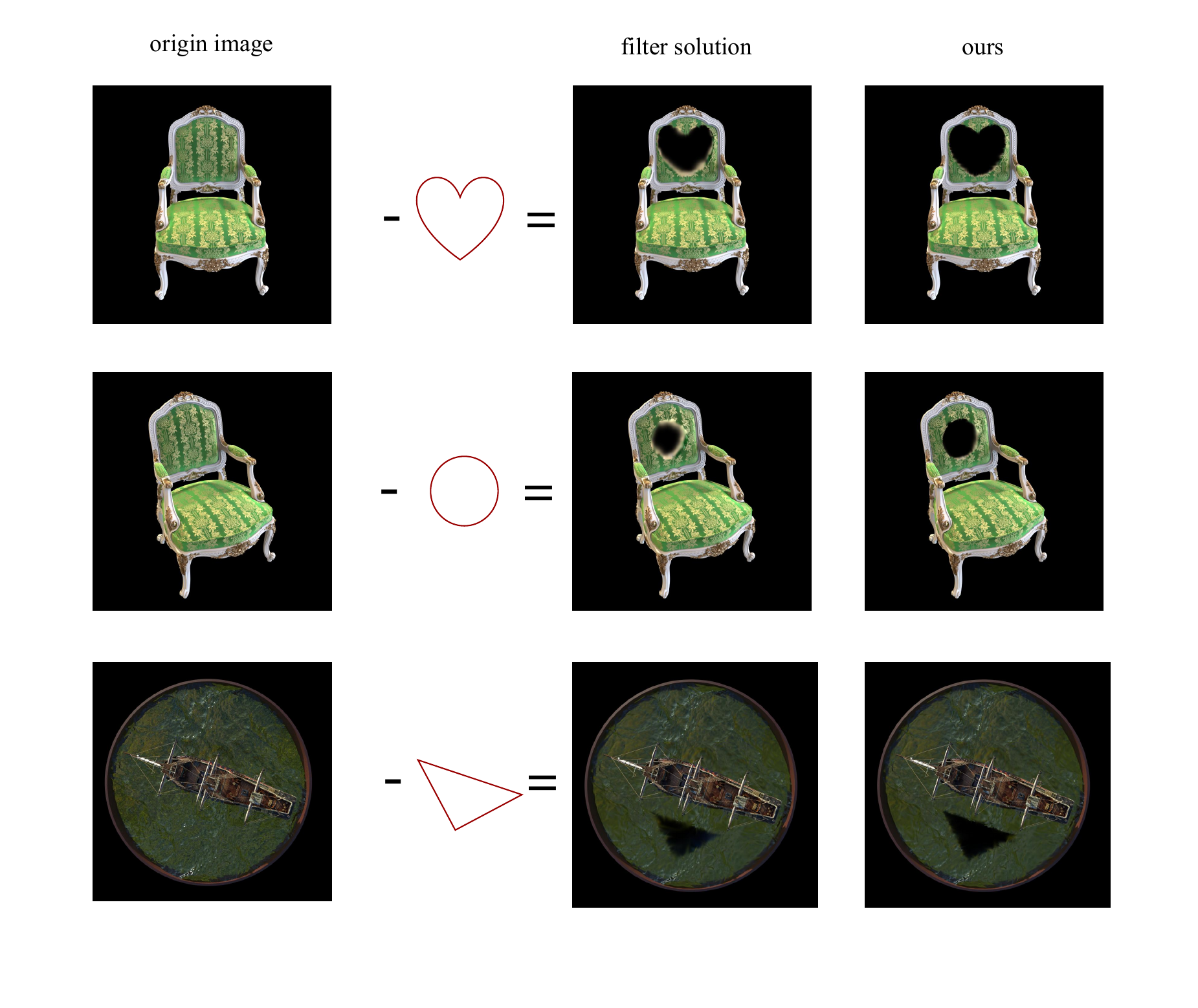}
    \caption{
    Various editing examples removing certain shapes from a 3D Gaussian splatting model. The filter solution of deleting the Gaussians within the shape directly produces more irregular boundaries than our methods, and leaves material in the hole left by removal. 
    }
    \label{figapp:edit-label}
\end{figure*}

\begin{figure*}
    \centering
    \includegraphics[width=\linewidth]{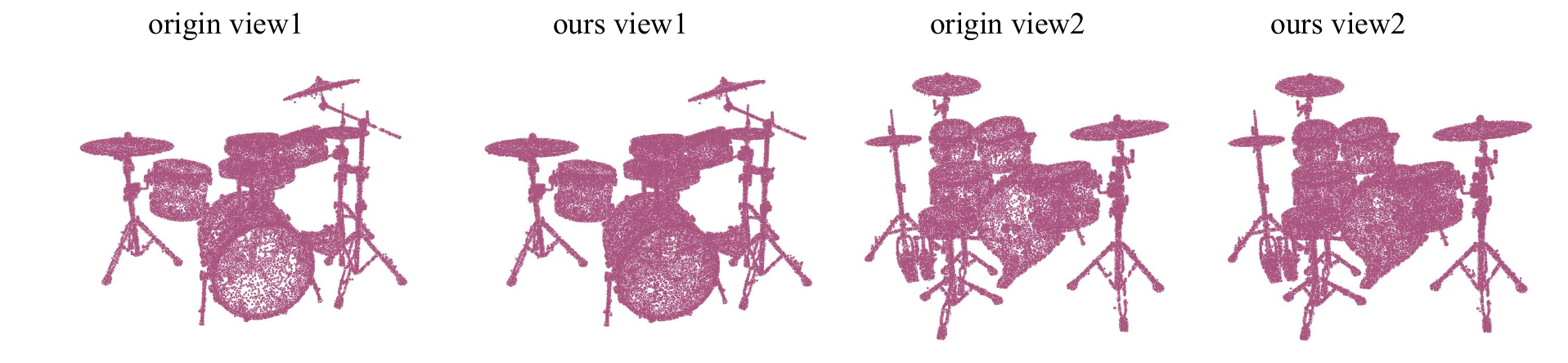}
    \caption{
    Example showing that our algorithm can help to extract a more uniform point cloud.
    }
    \label{figapp:refine-label}
\end{figure*}

\end{document}